\documentclass[12pt,preprint]{aastex}
\usepackage{float}

\newcommand{\axp}{\mbox{1E\,1841-045~}}
\newcommand{\axpnos}{\mbox{1E\,1841-045}}
\newcommand{\snr}{\mbox{Kes\,73~}}
\newcommand{\snrnos}{\mbox{Kes\,73}}

\shorttitle{}
\shortauthors{Lin et al.}

\begin{document}

\title{Burst and Persistent Emission Properties during the Recent Active Episode of the Anomalous X-ray Pulsar 1E\,1841-045}

\author{Lin Lin\altaffilmark{1, 2}, Chryssa Kouveliotou\altaffilmark{3}, Ersin G\"o\u{g}\"u\c{s}\altaffilmark{4}, Alexander J. van der Horst\altaffilmark{5}, Anna L. Watts\altaffilmark{6}, Matthew G. Baring\altaffilmark{7}, Yuki Kaneko\altaffilmark{4}, Ralph A.M.J. Wijers\altaffilmark{6}, Peter M. Woods\altaffilmark{8}, Scott Barthelmy\altaffilmark{9}, J. Michael Burgess\altaffilmark{2}, Vandiver Chaplin\altaffilmark{2}, Neil Gehrels\altaffilmark{9}, Adam Goldstein\altaffilmark{2}, Jonathan Granot\altaffilmark{10}, Sylvain Guiriec\altaffilmark{2}, Julie Mcenery\altaffilmark{9}, Robert D. Preece\altaffilmark{2}, David Tierney\altaffilmark{11}, Michiel van der Klis\altaffilmark{6}, Andreas von Kienlin\altaffilmark{12}, Shuang Nan Zhang\altaffilmark{1, 13}}
\email{lin198361@gmail.com}

\altaffiltext{1}{National Astronomical Observatories, Chinese Academy of Sciences, Beijing 100012, China}
\altaffiltext{2}{CSPAR, University of Alabama in Huntsville, Huntsville, AL 35805, USA}
\altaffiltext{3}{Space Science Office, VP62, NASA/Marshall Space Flight Center, Huntsville, AL 35812, USA}
\altaffiltext{4}{Sabanc\i University, Faculty of Engineering and Natural Sciences, Orhanl\i$-$Tuzla, \.{I}stanbul 34956, Turkey}
\altaffiltext{5}{Universities Space Research Association, NSSTC, Huntsville, AL 35805, USA}
\altaffiltext{6}{Astronomical Institute "Anton Pannekoek," University of Amsterdam, Postbus 94249, 1090 GE Amsterdam, The Netherlands}
\altaffiltext{7}{Department of Physics and Astronomy, Rice University, MS-108, P.O. Box 1892, Houston, TX 77251, USA}
\altaffiltext{8}{Corvid Technologies, 689 Discovery Drive, Huntsville, AL 35806, USA}
\altaffiltext{9}{NASA Goddard Space Flight Center, Greenbelt, MD 20771, USA}
\altaffiltext{10}{Centre for Astrophysics Research, University of Hertfordshire, College Lane, Hatfield, Herts, AL10 9AB, UK}
\altaffiltext{11}{University College, Dublin, Belfield, Stillorgan Road, Dublin 4, Ireland}
\altaffiltext{12}{Max Planck Institute for extraterrestrial Physics, Giessenbachstrasse, Postfach 1312, 85748 Garching, Germany}
\altaffiltext{13}{Key Laboratory of Particle Astrophysics, Institute of High Energy Physics, Chinese Academy of Sciences, P.O. Box 918-3, Beijing 100049, China}

\begin{abstract}

{\it Swift}/BAT detected the first burst from \axp in May 2010 with intermittent burst activity recorded through at least July 2011. Here we present {\it Swift} and {\it Fermi}/GBM observations of this burst activity and search for correlated changes to the persistent X-ray emission of the source. The $T_{90}$ durations of the bursts range between $18-140$\,ms, comparable to other magnetar burst durations, while the energy released in each burst ranges between $(0.8 - 25)\times10^{38}$ erg, which is in the low side of SGR bursts.  We find that the bursting activity did not have a significant effect on the persistent flux level of the source. We argue that the mechanism leading to this sporadic burst activity in \axp might not involve large scale restructuring (either crustal or magnetospheric) as seen in other magnetar sources.

\end{abstract}


\section{Introduction} 

Anomalous X-ray Pulsars (AXPs) form a small subset of slowly rotating neutron stars identified as a separate class by \citet{mereghetti1995} based on their persistent X-ray emission similarities that set them apart for the bulk of X-ray pulsars. Their spin periods, $P$, and spin-down rates, $\dot{P}$, fall within narrow ranges (2-12\,s and 5$\times10^{-13}-10^{-10}$ s/s, respectively). Their magnetic fields, estimated from $P, \dot{P}$ are in excess of $10^{14}$\,G, placing these sources in the group of magnetar candidates (neutron stars with extreme magnetic fields). While AXPs were identified from the properties of their persistent X-ray emission, the other members of this group, Soft Gamma Repeaters (SGRs), were discovered when they entered burst active periods, emitting multiple short, soft bursts \citep[see][for a review]{woods2006}. The first burst emission from an AXP was discovered in 2002 \citep{gavriil2002}. By now, bursts have been observed from almost all confirmed AXPs, convincingly linking these two types of neutron star \citep{mereghetti2008}.

Burst activity has been shown to affect the persistent emission and timing characteristics for almost all AXPs, while for SGRs such effects are consistently found only following energetic bursts \citep{woods2004,gavriil2004,gavriil2006,woods2007,israel2007,zhu2008,esposito2008,gogus2010,gonzalez2010,gogus2011a}. During the burst active period, the persistent X-ray emission of magnetars has been found to suddenly increase and then rapidly decrease according to an exponential decay that asymptotically approaches the pre-burst active level \citep{woods2006, rea2011}. The spectral and temporal properties of the emission also change during the outburst. For example, the X-ray flux of 1E\,$2259+586$ increased by at least a factor $\sim 20$ during the same time interval when more than 80 SGR-like bursts were emitted \citep{woods2004,gavriil2004}, and decayed steadily during the next three years to almost the preburst level \citep{zhu2008}. We report here on the unusual behaviour of the persistent X-ray emission of \axpnos, after its recent burst active period \citep{barthelmy2011}.

\axp was discovered in 1985 as an unresolved {\it Einstein} point source at the center of the \snr Supernova Remnant \citep[SNR;][]{kriss1985}. Later observations with the {\it Advanced Satellite for Cosmology and Astrophysics} ({\it ASCA}) revealed a period of $\sim 11.8$\,s \citep{vasisht1997}. This spin period was confirmed and a rapid secular spin-down rate of $\dot{P} = 4.16 \times 10^{-11}$\,s/s was derived with {\it Ginga}, {\it ASCA}, {\it Rossi X-Ray Timing Explorer} ({\it RXTE}), and {\it BeppoSAX} observations \citep{gotthelf1999, gotthelf2002}, corresponding to a dipole surface magnetic field of $\sim 7.1 \times 10^{14}$\,G. {\it Chandra} observations provided a precise location at R.A.$(J2000) = 18^{\rm h} 41^{\rm m} 19\fs343$, decl.$(J2000) = -04\arcdeg 56\arcmin 11\farcs16$ with a $1\sigma$ error of $0\farcs3$ \citep{wachter2004}. The source is on the Galactic Plane at a distance of $\sim 8.5_{-1.0}^{+1.3}$\,kpc \citep{tl2008} and with a large interstellar absorption preventing identification of an optical or infrared counterpart \citep{mereghetti2001,durant2005}. Unlike other magnetar candidates, \axp has a persistent X-ray emission which has remained constant for several decades \citep{gotthelf1999, zhu2010}.

Recently, \citet{kumar2010} reported the very first SGR-like burst from 1E\,1841-045, which triggered the Burst Alert Telescope onboard the {\it Swift} satellite ({\it Swift}/BAT) on 2010 May 6. They find that the burst was associated with a slight softening of the X-ray spectrum and a marginal ($\sim2\sigma$) increase in the persistent X-ray flux of the source. On 2011 February 8, the {\it Swift}/BAT detected another burst from \axp \citep{barthelmy2011}, but the {\it Swift}/X-ray Telescope ({\it Swift}/XRT) was unable to monitor the source as its direction was very close to the Sun \citep{barthelmy2011}. About 10 hours after the BAT trigger, the Gamma-ray Burst Monitor (GBM) onboard the {\it Fermi Gamma-ray Space Telescope} ({\it Fermi}) triggered on another short burst \citep{vdh2011} from the source direction. On 2011 February 9, the {\it RXTE} observed \axp for 3\,ks during which no additional bursts were detected. Moreover, the pulsed flux level did not change and there were no significant changes in the timing properties (i.e., offsets relative to the long-term rotational ephemeris) of the persistent emission \citep{gavriil2011a}. GBM detected two short and soft events with locations consistent with \axpnos, on 2011 February 17 \citep{tierney2011} and 21. We triggered a $\sim 4$\,ks {\it Swift}/XRT Target of Opportunity (ToO) observation on 2011 February 24 to monitor the source persistent X-ray emission. Additionally, to compare the post burst spectral state of the source with its historical behavior, we investigated ten earlier {\it Swift}/XRT observations with \axp in the field of view since 2008. During 2011 June 16 - July 2 there were 4 more events from \axpnos: two were detected with the {\it Swift}/BAT \citep{rowlinson2011,melandri2011}, and three with GBM. One event was detected with both instruments, namely the event on 2011 June 23.

In this Letter, we present our study with {\it Swift} and {\it Fermi}/GBM of the temporal and spectral properties of all nine bursts from \axpnos, and the evolution of the persistent emission of the source with {\it Swift}/XRT. Section \ref{data} describes the data reduction methods, and Section \ref{results} presents our results. We find that the spectral parameters and the unabsorbed flux of the persistent emission did not change significantly since 2008, even during the burst active period, and discuss the significance of our results in Section \ref{discussion}.

\section{Data reduction} \label{data}
\subsection{Swift Data}
We used the standard BAT software distributed within HEASoft v6.10 and the latest calibration files to process BAT data. First, we reran the BAT energy calibration task ({\it bateconvert}) to generate the detector quality map with bad and noisy detectors marked. We used the Bayesian blocks task {\it battblocks} to calculate the BAT burst durations (total time, $T_{90}$ and $T_{50}$\footnote{$T_{90}$ ($T_{50}$) are the times during which 90\% (50\%) of the burst counts are collected \citep{ck1993}}) with 2\,ms time resolution in the $15-150$\,keV energy range. We extracted 2\,ms binned, background subtracted, burst light curves in $15-150$\,keV with {\it batbinevt}. For the burst spectral analysis, we ran the mask weight task {\it batmaskwtevt} with the location of \axpnos. We extracted the standard 80 energy channel spectrum, integrated through the burst total durations, and updated the spectral keywords and the systematic errors. Finally, we generated the response matrix for the spectra with {\it batdrmgen}, and fit the time-integrated spectra with XSPEC v12.6.0.

There are 14 XRT observations of \axpnos, including our ToO. Of these only ten were in Photon Counting (PC) mode, providing the required spatial resolution to reliably extract source counts from the center of \snrnos. For these observations, we extracted the spectra of \axp from the Level 2 event data with the standard grade selection of $0-12$ in a circular region centered on the source location with a radius of $15\arcsec$. We selected the  background region carefully using the same radius of the source, within the \snr area (of radius $\sim 2\arcmin$) avoiding X-ray bright areas in \snrnos. We built the exposure map for each observation with {\it xrtexpomap}. Then we generated the ancillary response files (ARFs) with {\it xrtmkarf} for each spectrum with the point-spread function correction. Finally, we regrouped the \axp spectra with a minimum of 20 source counts per bin, and fit the resulting data in XSPEC v12.6.0 using the latest spectral redistribution matrix (RMF, {\it swxpc0to12s6\_20070901v011.rmf}).

\subsection{Fermi/GBM data}

\begin{figure}[h]
\includegraphics*[bb=20 0 490 340, scale=0.9]{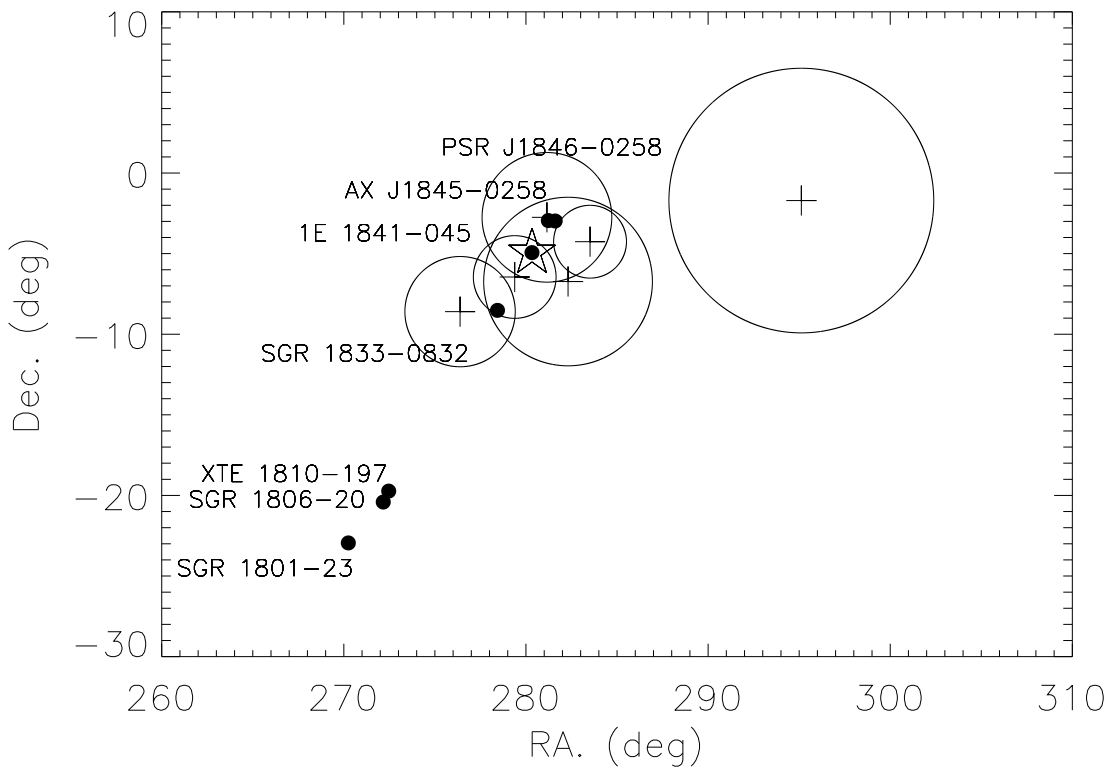}
\caption{Locations of the four BAT bursts ({\it star}) and 6 GBM bursts ({\it crosses}) with $1\sigma$ error circle. The {\it Chandra} location of \axpnos ({\it filled dot within the star}), and 6 other nearby magnetar candidates are also indicated ({\it filled dots}).} \label{location}
\end{figure}

The GBM locations of the \axp bursts have large statistical uncertainties, indicated by the $1\sigma$ error circles in Figure \ref{location}. Unfortunately, there are no simultaneous observations with other satellites that could narrow down these error circles, thus we cannot exclude the possibility that these bursts came from other known magnetars in the vicinity or even from a new source. However, the four {\it Swift}/BAT bursts are well localized ($\sim 1\arcmin$) at the \axp {\it Chandra} position. Since it is rare \citep[only twice before,][]{ibrahim2004,esposito2011} to have two different nearby magnetar sources emit bursts in the same time period within two weeks from each other, we conclude that it is reasonable to assume that the six GBM bursts are indeed from \axpnos. 

We selected the GBM NaI detectors \citep{meegan2009} with an angle to the source smaller than $50^{\circ}$ and not blocked by other parts of the satellite for all six bursts. We only used Time Tagged Event (TTE) data for our analyses, because of their fine temporal and spectral resolution \citep{meegan2009}. We also searched the entire February data and the interval between June 10 -- July 6 for untriggered bursts from \axp using the same algorithm described in \citet{kaneko2010}, and found one additional short burst on 2011 February 17 at 06:13:14 (UT) from the same general direction as \axpnos. Unfortunately, no TTE data were available for this untriggered burst, so it was not included in further analyses. We calculated the $T_{90}$ ($T_{50}$) durations for each burst in both count and photon space in $8-100$\,keV and in 2\,ms time bins \citep[for a detailed description see ][]{lin2011}. We generated the response files for each detector with the GBM response generator {\it gbmrsp v1.9} and analyzed the burst spectra ($8-200$\,keV) with the GBM public software tool RMFIT v3.3\footnote{http://fermi.gsfc.nasa.gov/ssc/data/analysis/user/} \citep[for a description of this tool see][]{kaneko2006}.

\section{Results}  \label{results}
\subsection{Burst Properties}
\begin{figure}[h]
\includegraphics*[bb=0 0 490 380, scale=0.3]{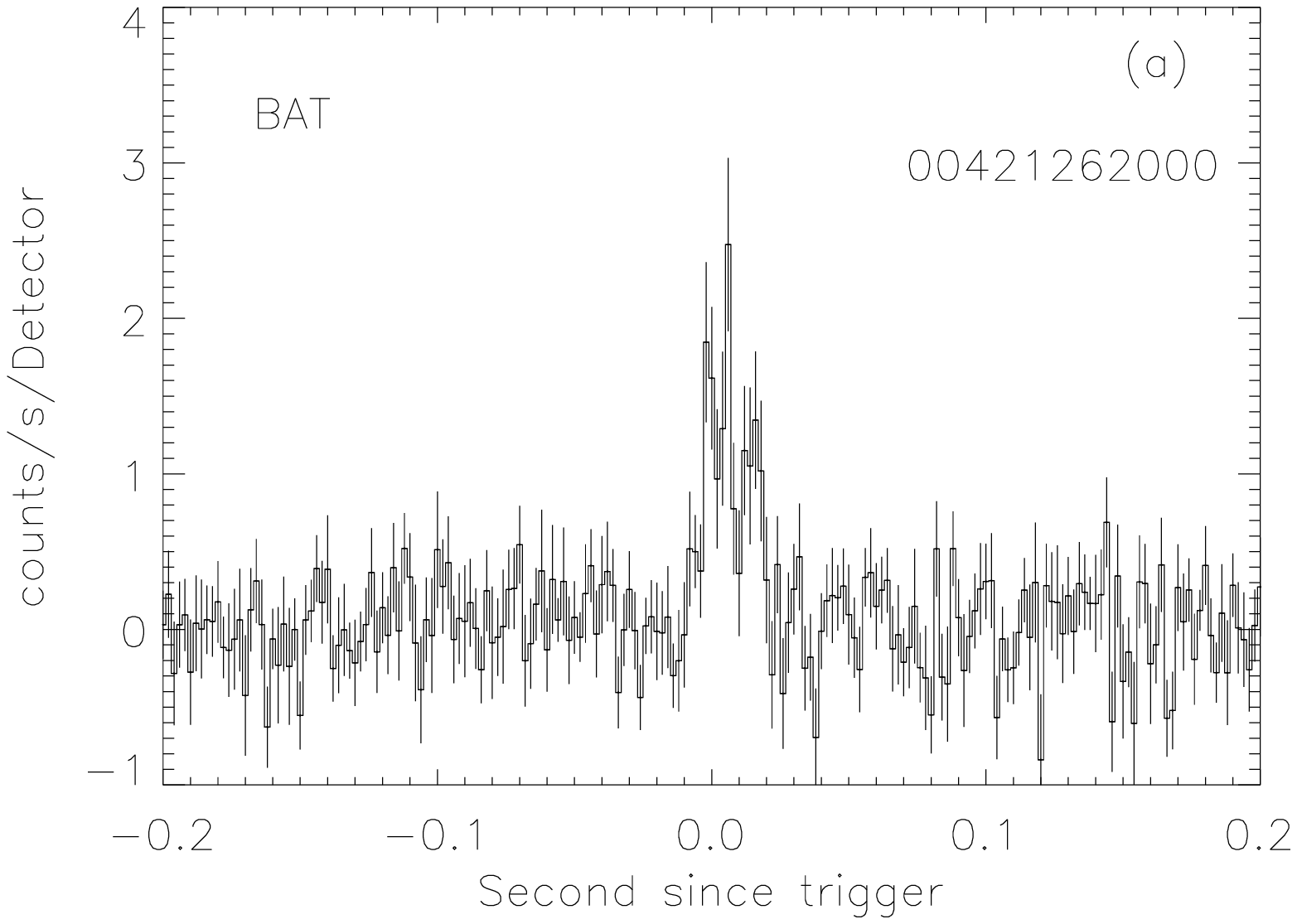}
\includegraphics*[bb=0 0 490 380, scale=0.3]{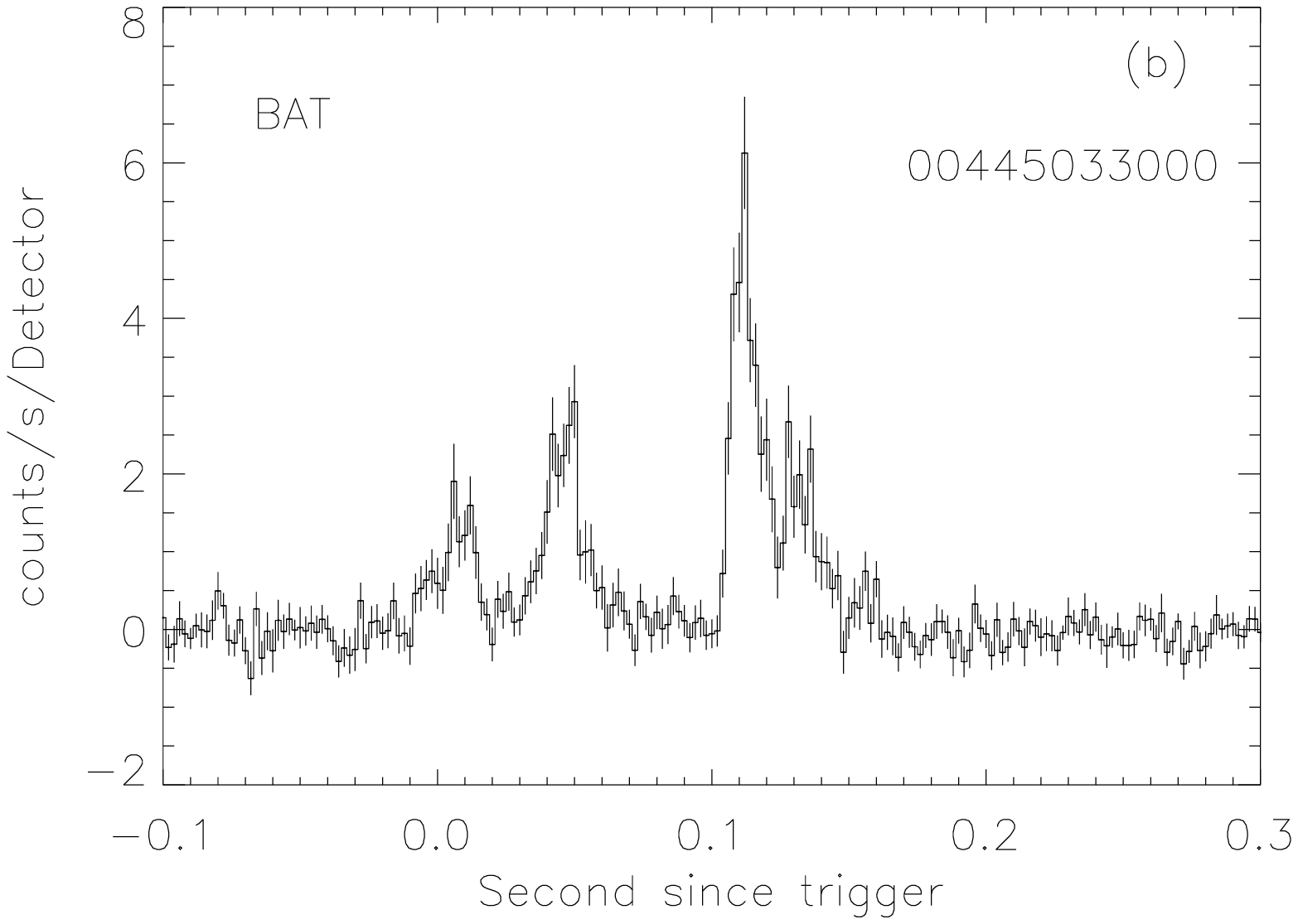}
\includegraphics*[bb=0 0 490 380, scale=0.3]{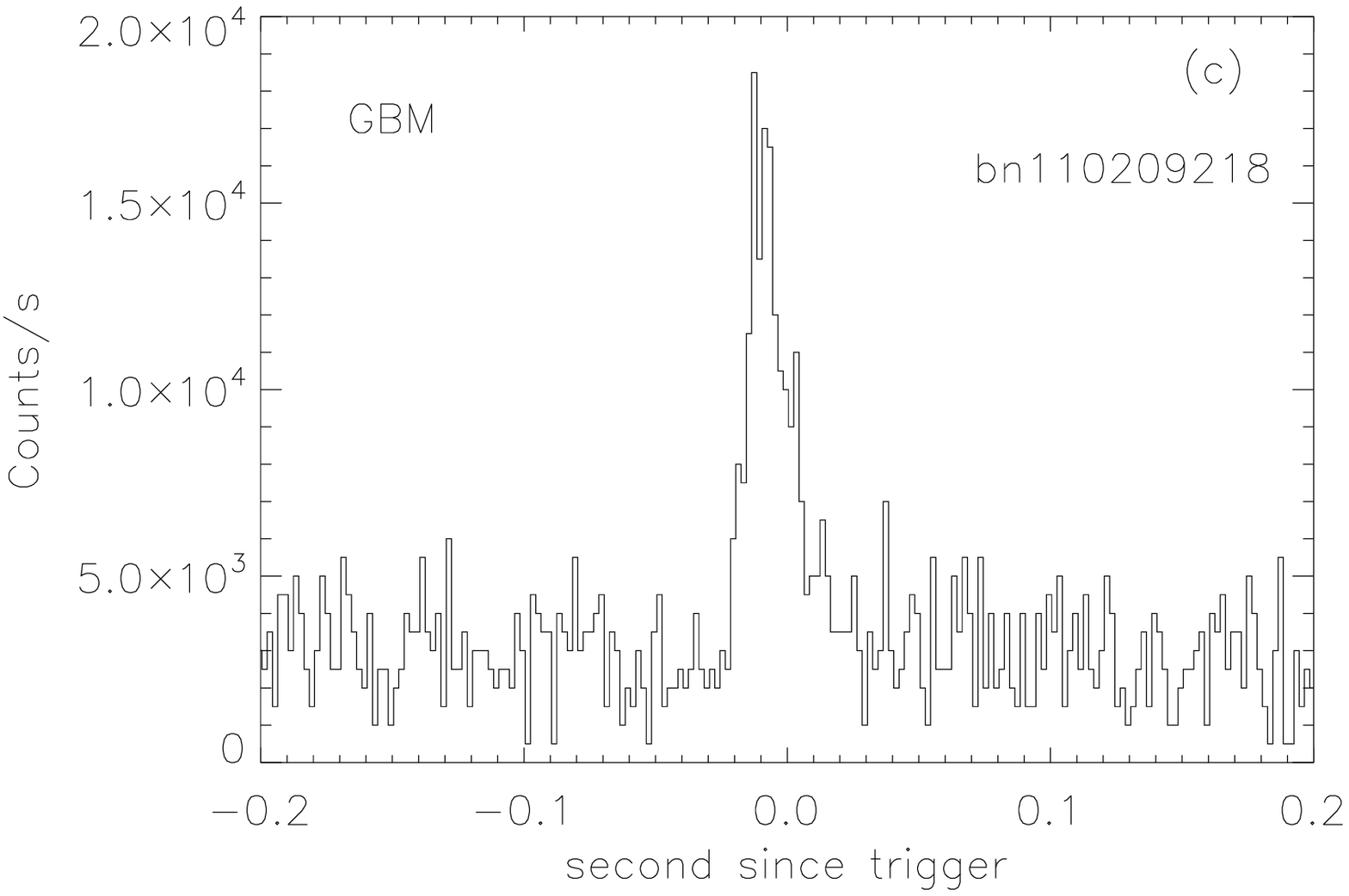}\\
\includegraphics*[bb=0 0 490 380, scale=0.3]{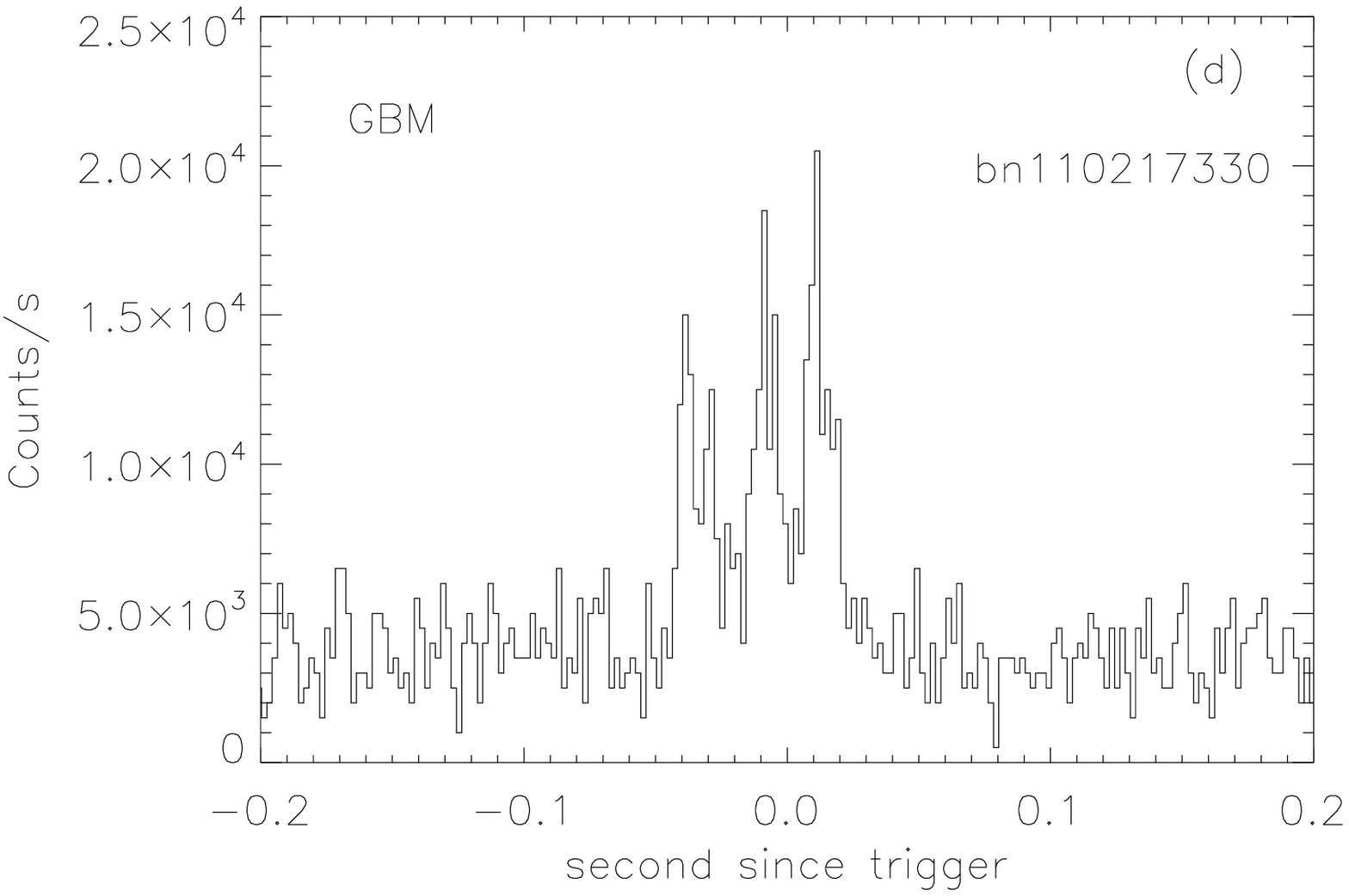}
\includegraphics*[bb=0 0 490 380, scale=0.3]{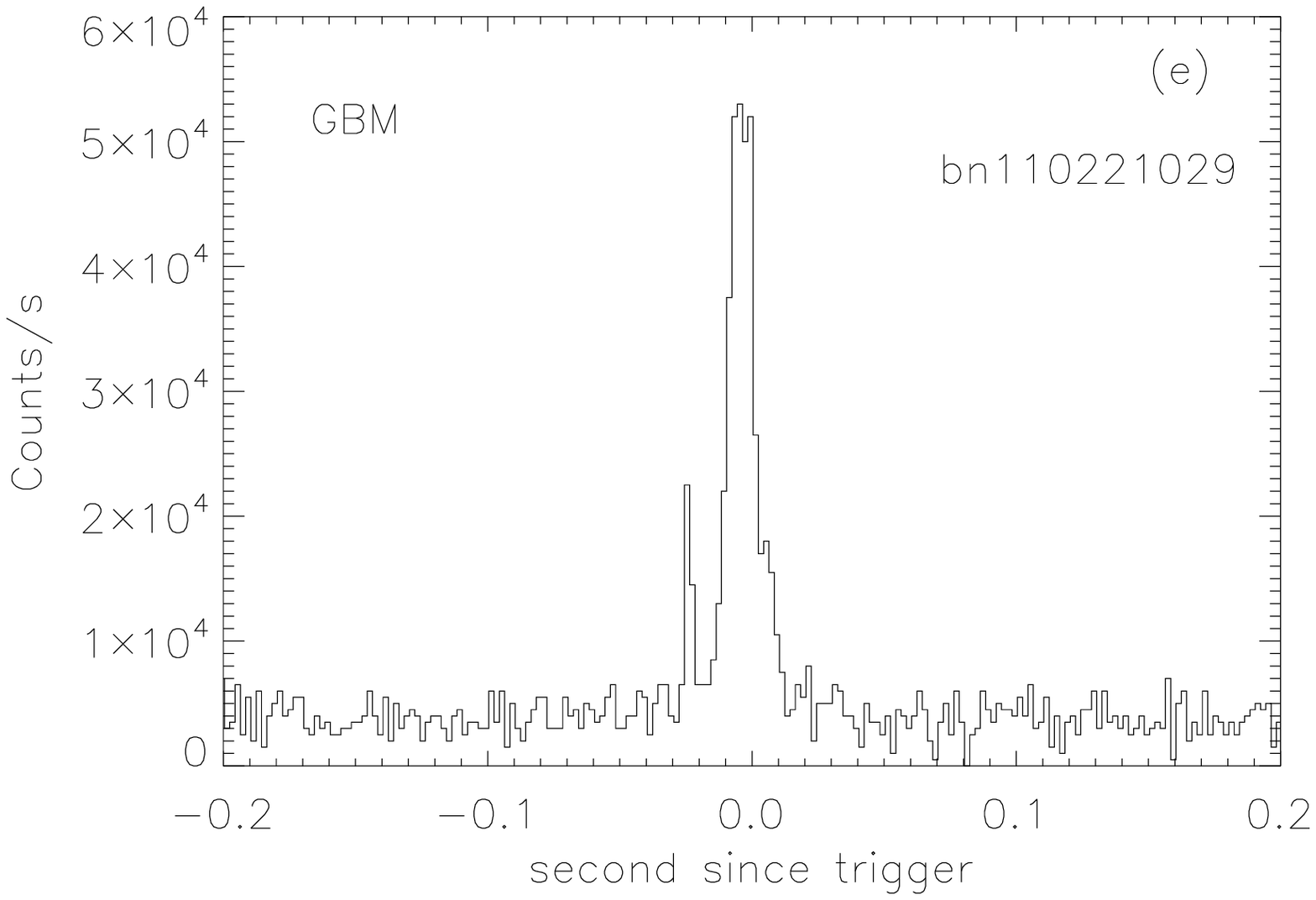}
\includegraphics*[bb=0 0 490 380, scale=0.3]{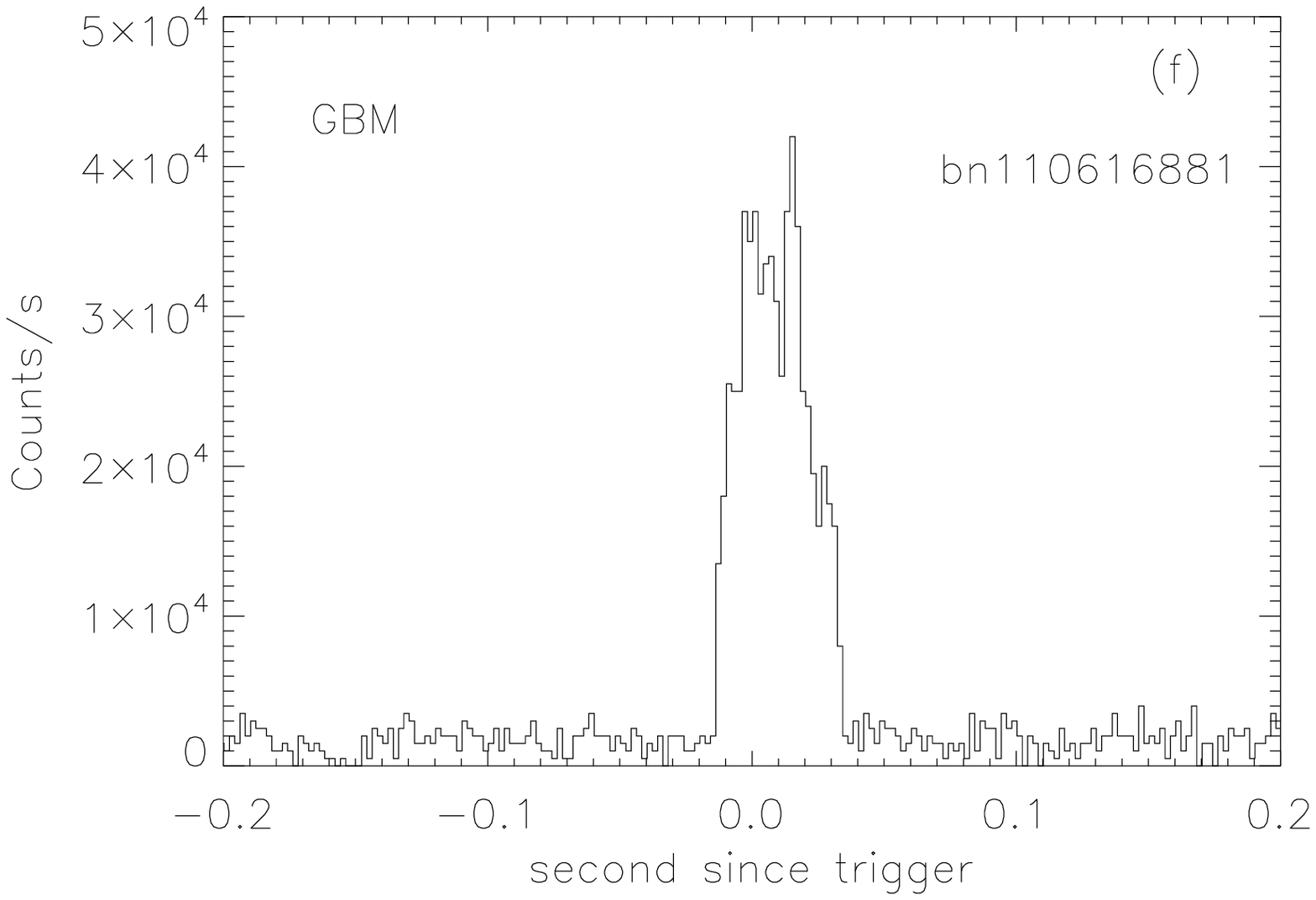}\\
\includegraphics*[bb=0 0 490 380, scale=0.3]{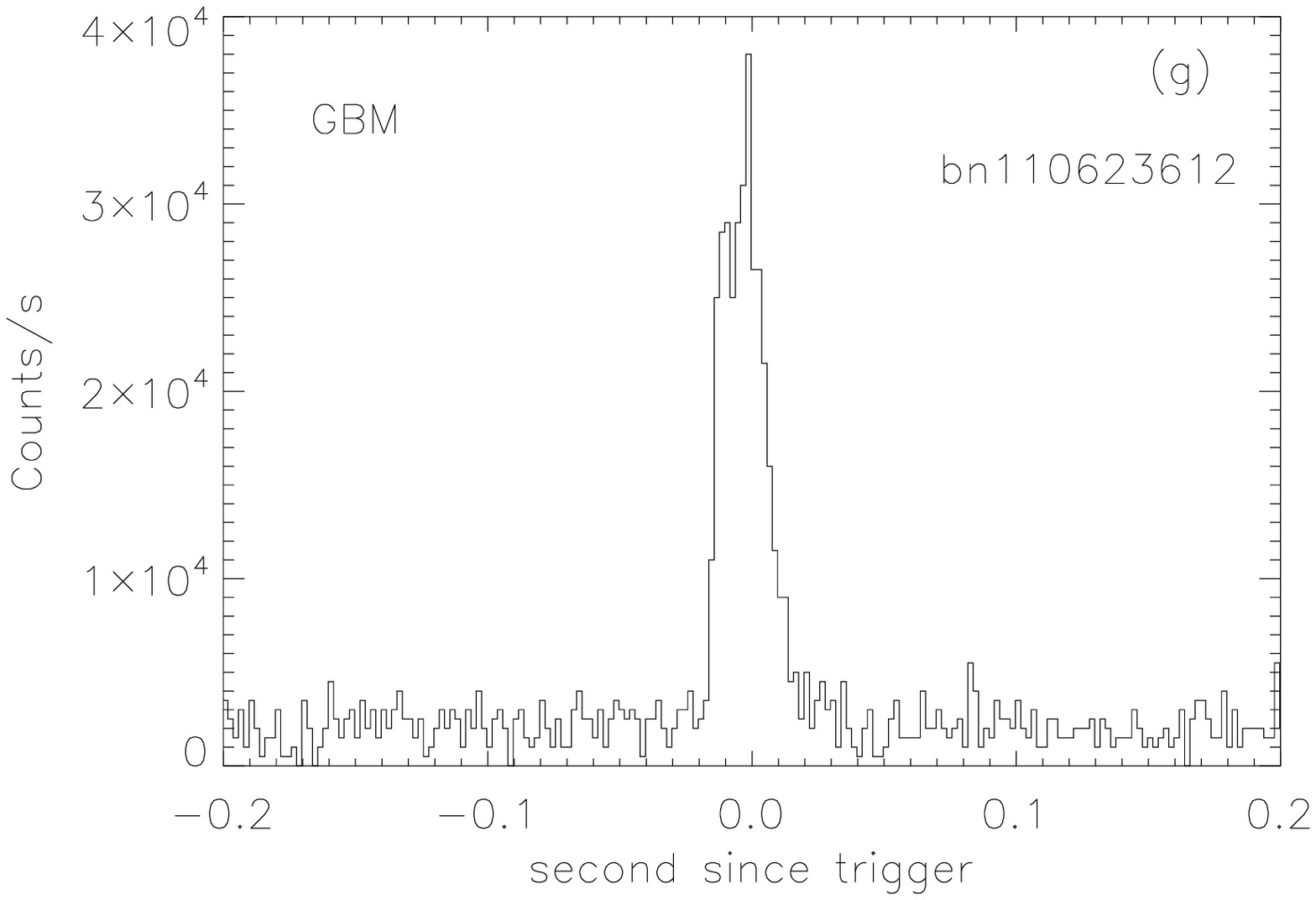}
\includegraphics*[bb=0 0 490 380, scale=0.3]{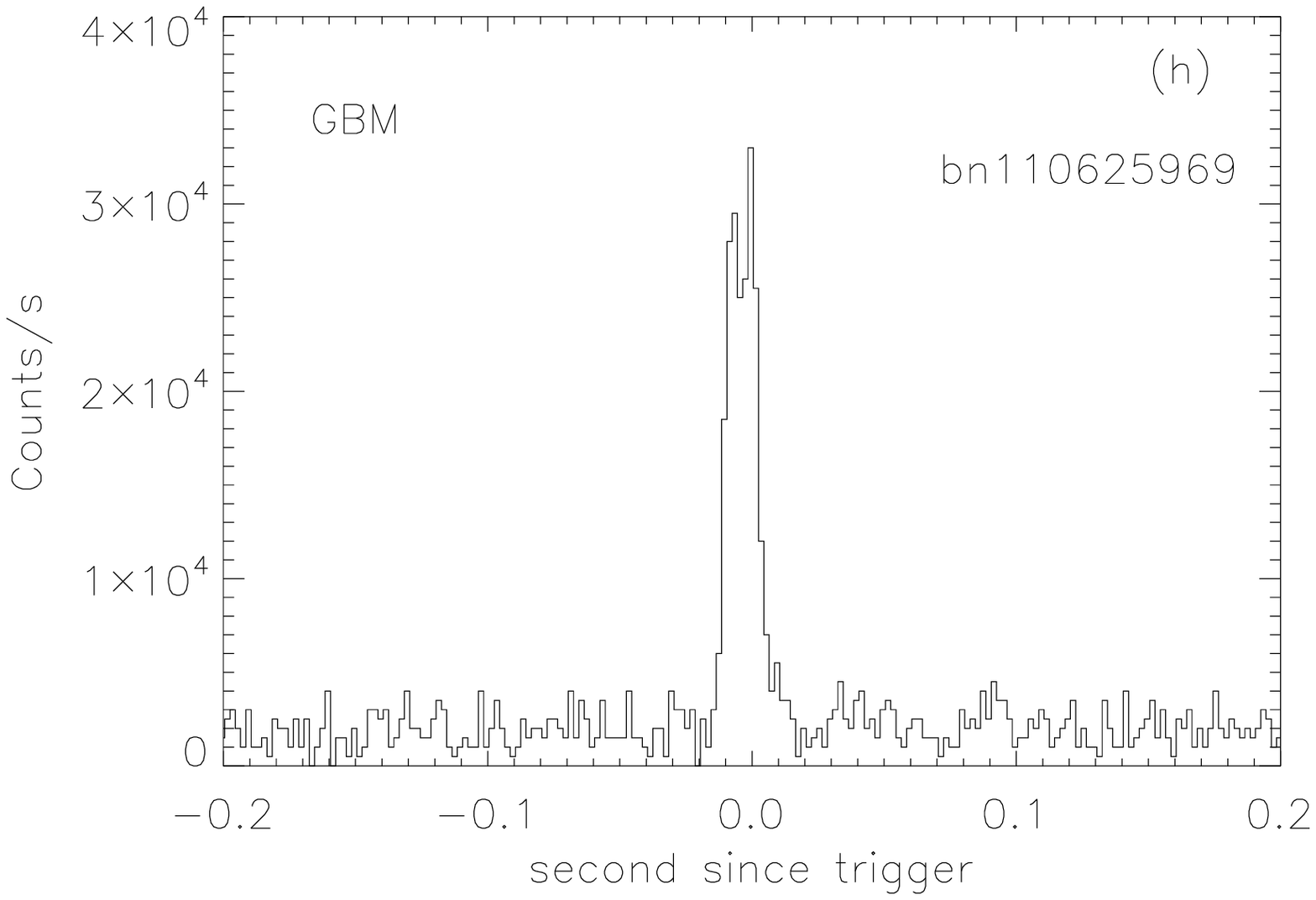}
\includegraphics*[bb=0 0 490 380, scale=0.3]{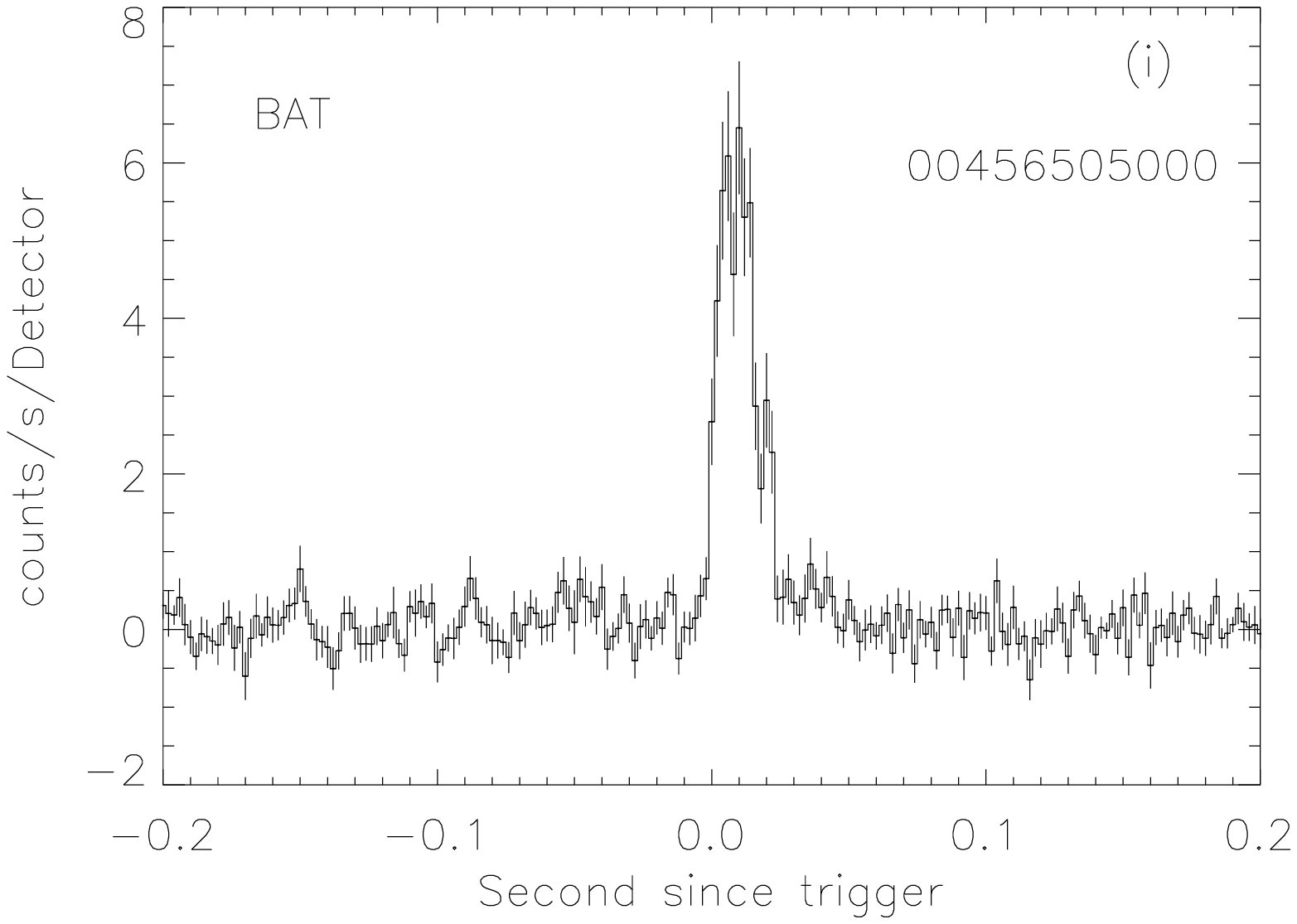}
\caption{({\it panels a, b and i}): Background subtracted 2\,ms time resolution light curves ($15-150$\,keV) for the three \axp bursts detected with {\it Swift}/BAT. ({\it panels c--h}): 2\,ms binned raw count rate light curves of six {\it Fermi}/GBM bursts from \axp ($8-100$\,keV). } \label{burstlcspec}
\end{figure}

We analyzed here for consistency, in addition to the February and June--July bursts from \axpnos, also the 2010 May 6 {\it Swift}/BAT trigger reported by \citet{kumar2010}. Figure \ref{burstlcspec} (a--i) exhibits the time profiles of all bursts; these are single or multi-peaked similar to other magnetar candidate bursts. We did not detect any thermal tail emission after the very bright burst in Figure \ref{burstlcspec}(e), as is often observed in bright AXP/SGR bursts \citep{lenters2003,gogus2011b}. The $T_{90}$ durations of the bursts range between $18-140$\,ms, comparable to other magnetar burst durations \citep{gogus2001,gavriil2004,lin2011}. Table \ref{burstobs} (columns $1-5$) lists the trigger date, trigger time, the selected NaI detectors (for GBM bursts only), and the durations of all nine bursts.

We fit several models to the burst spectra: a single power-law, Optically Thin Thermal Bremsstrahlung (OTTB), single Black Body (BB), a power law with an exponential cut-off (COMPT), and two BBs. We note that magnetar model motivations for multi-component BBs or Comptonization-type spectra mimicked by COMPT forms, are discussed in detail in \citet{lin2011}. Here we find that a COMPT model can fit all burst spectra except for the faintest one (Figure \ref{burstlcspec}(a)), where the COMPT model parameters can not all be constrained. We fit that burst spectrum with a single BB model, which has one parameter less. This event was analyzed earlier by \citet{kumar2010}, who fit the spectrum with three Gaussian functions. However, \citet{kumar2010} used the {\it Swift}/BAT location to create the response files for their spectral analysis, which placed the source roughly 1$\arcmin$ away from the accurate {\it Chandra} location used in the current analysis. Therefore, their background subtracted spectrum may have been contaminated by the contribution of the SNR Kes 73. This contribution cannot be removed with mask-weighting of the BAT events, and might have led to the appearance of unusual spectral lines in the spectrum.

The brightest burst (Figure \ref{burstlcspec}(e)) has enough statistics to also allow a fit using a two BB model; we used the Castor modified\footnote{heasarc.gsfc.nasa.gov/docs/xanadu/xspec/wstat.ps} Cash-statistic \citep{cash1979} (C-stat) to determine the goodness of fit for each model. This is a modified maximum likelihood estimator which asymptotes to $\chi^2$, used when there are small numbers of counts/bin (Poisson regime), which is the case for most of the SGR events (especially in the higher energy bins). The C-stat value for the two BB fit (293.8 for 298 dof) is similar to that of the COMPT model fit. The temperature of the hot and cool BB components are $13.1\pm1.2$\,keV and $5.6\pm1.1$\,keV, and the corresponding radii of the emitting areas are 2.1 and 7.9 km, respectively. We also performed a joined fit between BAT and GBM for the common event of 2011 June 23. The model parameters, statistics and burst energetics are listed in Table \ref{burstobs} (columns $6-10$). The fluences and $E_{\rm peak}$ values of the eight bursts that could be fit with the COMPT model range between $\sim 4\times10^{-8} - 2.9\times10^{-7}$\,erg cm$^{-2}$, and $\sim 28-51$\,keV, respectively. 

\subsection{Persistent emission light-curve}
\begin{figure}[h]
\includegraphics*[bb=20 0 500 350, scale=0.9]{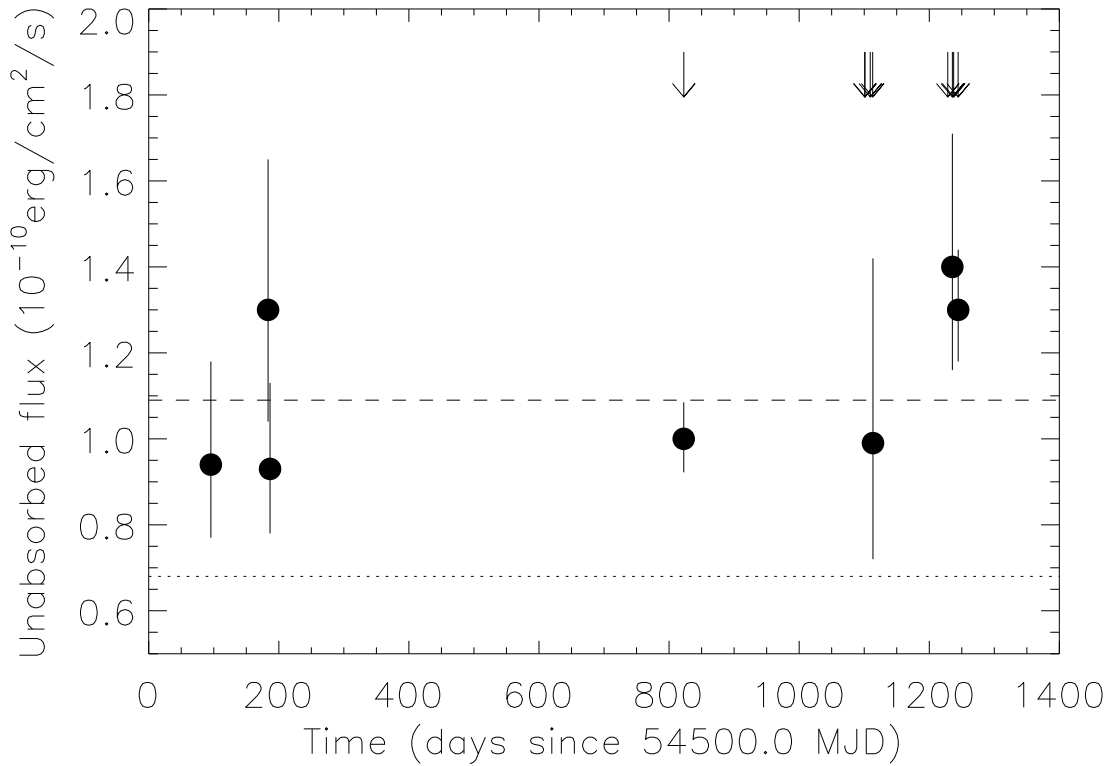}
\caption{The $0.5-10$\,keV unabsorbed flux of the XRT observations of the persistent emission of \axpnos. The arrows indicate the times of the burst emission. The dashed line is the weighted mean of the seven XRT datasets. The dotted line indicates the historical quiescent unabsorbed flux level from the {\it Chandra} observation \citep{morii2003}.} \label{persisflux} 
\end{figure}

We fit the spectra of the persistent emission from \axp with a single power-law model modified by interstellar absorption. When the separation between two XRT observations was very short, we combined the data to improve the statistics (e.g., after the 2010 observations). We noticed that the  $N_{\rm H}$ remained constant (within errors) in all fits. We then fit all XRT observations at the same time with linked $N_{\rm H}$, obtaining a value for the latter of $2.40_{-0.11}^{+0.12}\times10^{22}$\,cm$^{-2}$. Table \ref{persistantobs} lists the observation ID, observation date, exposure time, count rate, power-law index, statistics and unabsorbed flux in $0.5-10$\,keV for the data sets in PC mode used here. Figure \ref{persisflux} presents the time history of the unabsorbed flux in $0.5-10$\,keV of all observations. 

The unabsorbed flux level was first calculated in 1997 using the {\it ASCA} data by \citet{vasisht1997}, to be $6.3\times10^{-11}$\,erg cm$^{-2}$ s$^{-1}$ (within the same energy range and with the same model). Later \citet{morii2003} estimated a flux of $6.8\times10^{-11}$\,erg cm$^{-2}$ s$^{-1}$ from the source in 2000, using {\it Chandra} observations. Although the {\it Chandra} and {\it XMM-Newton} observations of \axp could be fit with two components \citep[absorbed BB+PL;][]{morii2003,kumar2010}, the XRT data could not constrain the parameters of a two-component fit. \citet{kumar2010} have also reached the same conclusion.

During the first 200 days of the \axp light curve shown in Figure \ref{persisflux}, the XRT flux measurements are compatible (within $2.0\sigma$) with the {\it Chandra} historical flux (Figure \ref{persisflux}, dotted line) value reported by \citet{morii2003}. After day 800 three of the four XRT measurements deviate between $3.0-5.0\sigma$ from this value, indicating a possible increase associated with the source burst activity. However, a power law fit of the entire XRT data set resulted in a positive slope of $0.11\pm0.07$, indicating an almost constant flux level during the 1400 day interval. Earlier, \citet{kumar2010} reported a marginal persistent flux increase in \axp associated with the SGR-like burst in 2010. We conclude that the current data are insufficient to significantly determine the trend of the persistent source emission. 

Finally, we estimated the (weighted) average unabsorbed flux of \axp to be $(10.9\pm0.6)\times10^{-11}$\,erg\,cm$^{-2}$\,s$^{-1}$ (Figure \ref{persisflux}, dashed line); at the source distance of $\sim8.5$\,kpc, the average isotropic persistent luminosity of the source is $(9.5\pm0.5)\times10^{35}$\,erg\,s$^{-1}$.

%Using the average unabsorbed flux of $\sim10.9\times10^{-11}$\,erg\,cm$^{-2}$\,s$^{-1}$ (Figure \ref{persisflux}, dashed line) and the source distance of $\sim8.5$\,kpc, we estimate the average isotropic persistent luminosity of \axp to be $\sim9.5\times10^{35}$\,erg\,s$^{-1}$. This luminosity, as well as the spectral parameters of the emission (see Table \ref{persistantobs}), are stable (within $\sim2\sigma$) from 2008 to 2011, despite the 2011 active burst periods. Earlier, \citet{kumar2010} reported a marginal persistent flux increase in \axp associated with the SGR-like burst in 2010. From Figure \ref{persisflux} we find that the 2010 flux level ($\sim$ day 800) is $<1\sigma$ away from  the average value of the persistent X-ray emission of the source (Figure \ref{persisflux}, dashed line).  Overall the persistent emission level seems to be constant within $1.5\sigma$ during the last 1400 days of source monitoring with the {\it Swift}/XRT. The 2011 burst activity thus seems to not have affected the level of the persistent source emission of \axpnos, at least during the period covered in Figure \ref{persisflux}. We note that the current average flux level is $\sim60\%$ above the historical quiescence level (Figure \ref{persisflux}, dotted line). However, this difference is roughly $0.5 - 2.0 \sigma$ significant. Finally,  we found no burst activity during the (apparently) increased flux level in 2008 (days 0-200) in the {\it RXTE} and GBM data. 

\section{Discussion} \label{discussion}

We have analyzed here all nine bursts from \axp detected during 2010/2011 with {\it Swift}/BAT and {\it Fermi}/GBM. We found that their spectral and temporal properties are quite similar to those of typical SGR bursts. The energy released in these bursts ranges between $(0.8 - 25)\times10^{38}$ erg, with a total of $\sim8\times10^{39}$ erg released in the eight bursts of 2011. Note that these energies are on the low side of SGR bursts \citep{woods2006}. Moreover, \axp is not an efficient burster: only four bursts were seen in its February 2011 active episode, and another four during 2011 June -- July, while prolific SGRs (e.g., SGR$1900+14$ or SGR 1806$-$20) can emit up to thousands of short bursts when active.  

One of our intriguing findings is that this low-level burst activity had very low impact on the source persistent emission level, in contrast to the changes associated with such activity observed in almost all AXPs in the past. We note here, however, that a prominent AXP, 4U$0142+01$, emitted six X-ray bursts in 2006 and in early 2007, but also showed no remarkable change in its persistent X-ray flux \citep{gonzalez2010,gavriil2011b}. An SGR persistent emission would typically not have been affected by these burst intensity levels \citep[see e.g.,][]{woods2007}.  Future XRT observations, in the absence of renewed burst activity, will determine whether the source flux will return to its historic  quiescent level.

Large changes in the persistent emission after intense bursting activity have been observed in other magnetar candidates, and are expected on theoretical grounds. Bursting activity is thought to be associated with the release of magnetic stress, triggered either by crust rupturing \citep{td1995} or magnetospheric instabilities \citep{lyutikov2003}. The result should be reconfiguration of the field geometry and/or scattering properties of the magnetosphere \citep{tlk2002,woods2001}, both of which should affect the flux and pulse shape of the persistent emission. Surface heating and enhanced thermal emission are also expected to result from crustal shear or impact of particles due to magnetic reconnection.

The lack of significant flux enhancement in conjunction with bursting in \axp may imply that the fracturing of the neutron star crust (or the magnetospheric instability) in \axp was not a large-scale one, and did not thus have any detectable impact on the longer lasting persistent source properties \citep[including its spin characteristics, see][]{gavriil2011a}. Another possibility is that the magnetospheric dissipation of the burst energy is largely directed away from the atmospheric zones that spawn the persistent emission.

\acknowledgments 
This publication is part of the GBM/Magnetar Key Project (NASA grant NNH07ZDA001-GLAST, PI: C. Kouveliotou). E.G. and Y.K. acknowledge the support from the Scientific and Technological Research Council of Turkey (T\"UB\.ITAK) through grant 109T755. ALW acknowledges support from a Netherlands Organisation for Scientific Research (NWO) Vidi Grant. MGB acknowledges support from NASA through grant NNX10AC59A. RAMJW acknowledges support from the European Research Council via Advanced Investigator Grant no. 247295. A.v.K. was supported by the Bundesministeriums f\"ur Wirtschaft und Technologie (BMWi) through DLR grant 50 OG 1101.

\begin{deluxetable}{lllccccccc}
\tabletypesize{\scriptsize}
\tablecaption{Bursts from \axp detected with {\it Swift}/BAT and {\it Fermi}/GBM. \label{burstobs}}
\tablewidth{0pt}
\tablehead{
\colhead{Date} & \colhead{Trigger time} & \colhead{NaI} & \colhead{$T_{90}$\tablenotemark{a}} & \colhead{$T_{50}$\tablenotemark{a}} & \colhead{Spectral\tablenotemark{b}} & \colhead{$E_{\rm peak}$\tablenotemark{b}} & \colhead{Stat/dof.\tablenotemark{b,c}} & \colhead{Fluence\tablenotemark{b}} & \colhead{$E_{\rm iso}$\tablenotemark{d}}\\
  & \colhead{UT} & \colhead{Det.} & \colhead{(ms)} & \colhead{(ms)} & \colhead{Index} & \colhead{(keV)} &  & \colhead{($10^{-8}$erg cm$^{-2}$)} & \colhead{($10^{38}$erg)}
}
\startdata
$10/05/06$ & 14:37:44.899 & - & $20 \pm 4$ & $12 \pm 4$ & - & $9.2_{-0.9}^{+0.8}$ \tablenotemark{e} & $53.01/56$ & $0.88_{-0.27}^{+0.04}$ & $0.76$ \\
$11/02/08$ & 19:17:27.739 & - & $136 \pm 18$ & $76 \pm 6$ & $0.34_{-0.44}^{+0.49}$ & $40\pm2$ &$44.44/55$ & $7.5_{-0.6}^{+0.2}$ & $6.5$ \\
$11/02/09$ & 05:14:25.944 & 0, 1, 2 & $36^{+22}_{-4}$ & $10 \pm 2$ & $-0.19_{-0.41}^{+0.45}$ & $51_{-4}^{+5}$ & $201.06/201$ & $5.1 \pm 0.4$ & 4.4 \\
$11/02/17$ & 07:55:55.295 & 0, 1, 6, 9 & $76^{+88}_{-16}$ & $42 \pm 4$ & $0.44_{-0.40}^{+0.44}$ & $45_{-2}^{+3}$  & $305.07/270$ & $8.4 \pm 0.5$ & 7.3 \\
$11/02/21$ & 00:41:16.252 & 0, 1, 2, 5 & $30^{+16}_{-8}$ & $14 \pm 4$ & $0.11_{-0.27}^{+0.29}$ & $41\pm2$ & $294.24/269$ & $10 \pm 1$ & 8.7 \\
$11/06/16$ & 21:09:08.430 & 10, 11 & $42 \pm 4$ & $20 \pm 2$ & $-0.90_{-0.20}^{+0.20}$ & $28\pm2$ & $158.09/130$ & $29 \pm 1$ & 25\\
$11/06/23$ & 14:41:42.764 & 8, 11 & $26^{+16}_{-4}$ & $12 \pm 4$ & $-0.11_{-0.26}^{+0.27}$ & $40\pm2$ & $130.67/133$ & $19 \pm 1$ & 16\\
$11/06/23$\tablenotemark{f} & 14:41:42.674 & - & $20 \pm 4$ & $10 \pm 2$ & $0.40_{-0.74}^{+0.83}$ & $28_{-4}^{+3}$ & $44.55/55$ & $12_{-2}^{+1}$ & 10\\
$11/06/23$\tablenotemark{g} & BAT-GBM & - & - & - & $0.14_{-0.35}^{+0.38}$ & $37\pm2$ & $100.01/189$ & $17 \pm 1$ & 15\\
$11/06/25$ & 23:16:03.175 & 9, 11 & $18^{+10}_{-4}$ & $8 \pm 4$ & $-0.04_{-0.35}^{+0.37}$ & $37 \pm 2$ & $107.67/132$ & $11 \pm 1$ & 9.5\\
$11/07/02$ & 08:38:38.760 & - & $32 \pm 12$ & $10 \pm 3$ & $0.44_{-0.53}^{+0.59}$ & $34\pm2$ & $44.42/55$ & $4.3_{-0.4}^{+0.2}$ & 3.7\\
\enddata \\
\begin{flushleft}
$^a$ Count durations calculated in $8-100$\,keV (GBM) and $15-150$\,keV (BAT). \\
$^b$ Calculated with the COMPT model in $8-200$\,keV (GBM) and $15-150$\,keV (BAT), with $1\sigma$ error. \\
$^c$ C-stat for GBM data and $\chi^2$ for BAT data. \\
$^d$ Corresponding energy released isotropically in the $15-150$\,keV range, assuming an 8.5 kpc distance for \axpnos. \\
$^e$ The temperature of the single BB model.\\
$^f$ Also detected with \textit{Swift}/BAT; the observation ID is 00455904000. \\
$^g$ Joined fit between BAT and GBM data.
\end{flushleft}
\end{deluxetable}

\begin{deluxetable}{llcccccc}
\tabletypesize{\scriptsize}
\tablecaption{Persistent emission from \axp observed with {\it Swift}/XRT in PC mode. \label{persistantobs}}
\tablewidth{0pt}
\tablehead{
\colhead{ID} & \colhead{date} & \colhead{expo time} & \colhead{count rate\tablenotemark{a}} & \colhead{index} & \colhead{$\chi^2/dof$} & \colhead{unabsorbed flux\tablenotemark{b}}\\
 & & (s) & (cts/s) & & &\\
}
\startdata
00090026002 & 2008 May 09 & 4032 & $0.23 \pm 0.01$ & $2.9_{-0.2}^{+0.2}$ & 42.17/42 & $9.4_{-1.7}^{+2.4}$\\
00090026003 & 2008 Aug. 05 & 687 & $0.21 \pm 0.02$  & $3.1 \pm 0.3$ & 3.25/5 & $13_{-3}^{+4}$\\
00090026004 & 2008 Aug. 08 & 5498 & $0.22 \pm 0.01$ & $2.9 \pm 0.2$ & 40.4/54 & $9.3_{-1.5}^{+2.0}$\\
00421262000 \& 00421262002 & 2010 May 06 & 4821 & $0.19 \pm 0.01$ & $2.9 \pm 0.1$ & 57.15/41 & $10 \pm 1$\\
00445776000 \& 00031863005 & 2011 Feb. 18/24 & 2792 & $0.23 \pm 0.01$ & $2.9 \pm 0.1$ & 47.87/28 & $9.9_{-2.7}^{+4.3}$ \\
00455904000 & 2011 Jun. 23 & 624 & $0.26 \pm 0.02$ & $3.1 \pm 0.2$ & 3.06/6 & $14_{-2}^{+3}$\\
00456505000 \& 00456505001 & 2011 Jul. 02 & 3312 & $0.20 \pm 0.01$ & $3.0 \pm 0.1$ & 40.73/29 & $13 \pm 1$
\enddata \\
\begin{flushleft}
$^a$ Background subtracted. \\
$^b$ $0.5-10$\,keV in $10^{-11}$ erg cm$^{-2}$ s$^{-1}$. \\
\end{flushleft}
\end{deluxetable}

\end{document}